\documentclass[twocolumn,twocolappendix,tighten,]{aastex631}
\usepackage{amssymb, amsmath,framed}

\usepackage{bm}
\expandafter\ifx\csname package@font\endcsname\relax\else
 \expandafter\expandafter
 \expandafter\usepackage
 \expandafter\expandafter
 \expandafter{\csname package@font\endcsname}
\fi
\def\bq{\begin{equation}}
\def\eq{\end{equation}}
\def\bqy{\begin{eqnarray}}
\def\eqy{\end{eqnarray}}

\begin{document}
\title{Technosignatures longevity and Lindy's law
}

\author{Amedeo Balbi}
\affiliation{Dipartimento di Fisica, Universit\`a di Roma Tor Vergata, Via della Ricerca Scientifica, 00133 Roma, Italy}
\author{Claudio Grimaldi}
\affiliation{Laboratory of Statistical Biophysics, Ecole Polytechnique F\'ed\'erale de Lausanne - EPFL, 1015 Lausanne, Switzerland}
\affiliation{Centro Studi e Ricerche Enrico Fermi, 00184 Roma, Italy}

\begin{abstract}
The probability of detecting technosignatures (i.e.\ evidence of technological activity beyond Earth) increases with their longevity, or the time interval over which they manifest. Therefore, the assumed distribution of longevities has some bearing on the chances of success of technosignature searches, as well as on the inferred age of technosignatures following a first contact. Here, we investigate the possibility that the longevity of technosignatures conforms to the so-called Lindy's law, whereby, at any time, their remaining life expectancy is roughly proportional to their age. We show that, if Lindy's law applies, the general tenet that the first detected technosignature ought to be very long lived may be overruled. We conclude by discussing the
number of emitters that had to appear, over the history of the Galaxy, in order for one of them to be detectable today from Earth.
\end{abstract}

\section{Introduction}
\label{SecIntro}

In the modern search for extraterrestrial intelligence (SETI), the term ``technosignature" refers to any remotely detectable evidence of technological activity \citep{Tarter2006, Socas-Navarro2021}. This is not limited to the fingerprints of (intentional or unintentional) interstellar communication that have constituted the goal of traditional radio \citep{Margot2019} and optical \citep{Schwartz1961, Wright2018} searches, but can include many other possible evidences. For example, the infrared glow emitted by Dyson spheres or other astroengineering megastructures harvesting the energy of stars \citep{Dyson1960, Wright2020}, the signature of stellar transit from such structures \citep{Wright2019} or from artificial satellites \citep{Socas-Navarro2018}, and the electromagnetic radiation associated to such technological activities as asteroid mining \citep{Forgan2011}, artificial illumination on planetary night-sides \citep{Lingam2017}, atmospheric pollutants  \citep{Lin2014, Kopparapu2021, Haqq-Misra2022a, Haqq-Misra2022b}, interstellar propulsion \citep{Lingam2017a}, leftover artifacts \citep{Haqq-Misra2012,Davies2013}, etc.

It has long been recognized that the probability of detecting a technosignature depends on its longevity, $L$, that is, the timespan over which it manifests; in essence, the probability of detection grows with $L$, because long-lasting technosignatures are visible over a larger volume
of space than short-lived ones, and thus have a higher chance of being intercepted from our location at the present epoch \citep{Grimaldi2017, Balbi2018}. Since the event of detection weighs more favorably technosignatures with large $L$, it has been recently proposed that first contact will be with older
civilizations than our own \citep{Kipping2020}. More generally, the longevity of detected technosignatures could exceed millions of years, as it is  decoupled from both the epoch of their first appearance and the duration of the civilization that produced them \citep{Balbi2021}.

Of course, inferring the longevity of a detected technosignature relies on the underlying probability distribution function (PDF) of $L$, denoted 
$\rho_L(L)$: this is independent of whether a technosignature is detected or not. On the other hand, $L$ is an unknown parameter that has been the subject of much speculation since the early days of SETI. The difficulty lies mainly in the lack of an \textit{a priori} understanding of what would be the typical duration of technosignatures, even at the order-of-magnitude level. Actually, it is not even clear that the notion of ``typical'' can be applied, since the possible values of $L$ could be  so widely distributed that their average may not be descriptive at all. Because of such lack of knowledge about the
underlying distribution of $L$, $\rho_L(L)$ is usually taken to be as uninformative as possible.
For example, \citet{Balbi2021} used a log-uniform distribution of $L$ ($\rho_L(L)\propto 1/L$) which reflects the lack of information on $L$ even at the order-of-magnitude level, whereas \citet{Kipping2020} derived a uninformative PDF of the longevity by adopting an objective Jeffrey's prior for 
the shape parameter of an exponential PDF of $L$.

However, a complete agnosticism on the prior distribution of $L$ may not always be justified. For example, it can be argued that, for a broad class of technosignatures, longevity is necessarily limited by the energy required for their operation. We term this subclass of technosignatures {\em technoemissions}, as they imply some sort of activity necessary for their operation and maintainance, as opposed to ``passive'' technosignatures (such as various forms of relic artifacts,  pollutants, etc.) that could in principle last for a long time without direct intervention. An   
obvious example of this sort are radio transmitters, which, for a given power required for interstellar transmission, would consume an amount of energy at least proportional to $L$ (here we take in consideration also the energy required for the maintenance of the transmitter and of all the infrastructures
required for its functioning). High-gain antennas or intermittent isotropic transmitters require less power \citep{Benford2010a, Gray2020}, but the energy used still 
grows with longevity. Dysonian megastructures may in principle radiate for millions of years or more, but their functioning requires nevertheless 
that energy is spent in maintenance over such long periods of time \citep[even though they may still be remotely detectable long after the extinction of 
their constructors; see][]{Cirkovic2019}.

These considerations suggest that, for technosignatures meeting the aforementioned condition (specifically, energy consumption increasing with longevity in some unspecified way), we should not expect long-lived ones to outnumber or be as common as short-lived ones. The rationale here is that long-lasting technoemissions will be disfavored because of high energy demands, and their production will require to overcome substantial technological challenges (for example by improving efficiency, by exploiting new sources of energy or new physics discoveries).

While we do not expect all extraterrestrial species to have overcome such technological challenges, those that have successfully done
so and managed to generate technoemissions for a long time will likely be able to produce them for a long time in the future as well.

The latter argument suggests that the probability distribution of longevities can be shaped by the so-called ``Lindy effect", or ``Lindy's law" \citep{Mandelbrot1984,Taleb2012,Eliazar2017,ord2023lindy}: this is a common (albeit counterintuitive) phenomenon observed in technology, whereby the future life expectancy of an innovation is proportional to its current age, leading to a longevity that increases with time. In other words, technologies that have been proved to function for a long time are expected to also have a long projected duration (roughly proportional to their current age), and outlive newly adopted ones. This generates a power-law, fat-tailed  distribution of $L$ of the form: 
\begin{equation}
\label{power-law}
\rho_L(L)\propto L^{-\alpha},
\end{equation}
where the positive definite exponent $\alpha$ weights the population of long-lasting technoemissions compared to short-lived ones. 
For example, $\alpha=1$ reduces Eq.~\eqref{power-law} to the uninformative log-uniform PDF mentioned above, which claims complete 
ignorance about even the order of magnitude of $L$, whereas $\alpha>1$ implies that longer longevities are disfavoured 
over to shorter ones, in line with our argument that $L$ is constrained by energy demands.

Of course, in shaping the tecnoemission longevities in terms of a Lindy process we are implicitly assuming that alien technologies follow similar evolutionary paths that terrestrial ones. However, as pointed out by \cite{ord2023lindy}, the Lindy effect arises from a wide variety of mechanisms, and can provide useful epistemic guidance in cases when little information on lifespans exists.

Our main purpose here is to study how the adoption of such power-law models for $\rho_L(L)$ reflects on the inferred longevity of technosignatures after a detection. Our focus on power-law probability distributions also stems from the observation that
they are found to govern a variety of situations involving ranking, such as city population, number of citations, word frequency, earthquake magnitude, wealth distribution, and so on \citep[see, e.g.,][]{Newman2005}. In addition, as we will see, power-law distributions are flexible enough to model a wide range of behaviors for $L$. 

Next, we describe in more detail our statistical model and explore its consequences. 

\section{The model}
\label{sec:model}

In formulating our model, we assume that technoemissions are generated 
by statistically independent sources located in the Galaxy at random positions $\mathbf{r}$ with respect to the galactic center.
In addition, we adopt the rather reasonable hypothesis that the generation rate of technoemissions is independent of the spatial 
distribution of the emitters, so that the expected number of technoemissions per unit time and volume can be factorized 
as $\Gamma(t)\rho_E(\mathbf{r})$, where $\Gamma(t)$ gives the technoemission rate at a time $t$ before present
and $\rho_E(\mathbf{r})$ is the PDF of the emitter position, defined such that 
$\rho_E(\mathbf{r})d\mathbf{r}$ is the probability of an emitter being within the element volume $d\mathbf{r}$ centered around $\mathbf{r}$.
Here, we assume that $\rho_E(\mathbf{r})$ is proportional to the distribution of stars in the galactic thin disk and adopt the following 
axisymmetric PDF \citep{Grimaldi2017,Grimaldi2018}:
\begin{equation}
\label{rhoE}
\rho_E(\mathbf{r})=\frac{\exp(-r/r_s)\exp(-\vert z\vert/z_s)}{4\pi r_s^2 z_s}
\end{equation}
where $r$ is the radial distance from the galactic center, $z$ is the height from the galactic plane, $r_s =8.15$ kly, and $z_s =0.52$ kly.

Since our analysis will focus on the temporal dependence of $\Gamma(t)$ rather than on its absolute value, we introduce the
(unconditional) PDF of $t$ defined as $\rho_t(t)=\Gamma(t)/\bar{N}$, where $\bar{N}=\int_0^{T_G}\!dt\,\Gamma(t)$ gives the average 
number of emitters ever existed since the birth of the Galaxy $T_G\sim 10^{10}$ yr ago. Here, we mainly adopt two models for
$\rho_t(t)$: a PDF uniform in $t$, which corresponds to assuming a stationary rate in the interval $0\leq t\leq T_G$, and a PDF that
is uniform in the logarithm of $t$, $\rho_t(t)\propto 1/t$, in the interval $t\in[T_\textrm{min},T_G]$, where $T_\textrm{min}\neq 0$ 
makes the log-uniform prior integrable. Since $T_\textrm{min}$ is not expected to be smaller than the travel time of a photon from the
nearest extrasolar system, $\approx 1$ yr, in the following we take $T_\textrm{min}= 1$ yr.

As already pointed out in previous studies \citep{Grimaldi2017,Balbi2018,Grimaldi2018,Balbi2021}, a necessary condition for any 
technoemission to be potentially detectable is that it must be located within our past light cone, that is, for a signal generated in 
$\mathbf{r}$ at time $t$ to be observable, the following condition must be fulfilled:
\begin{equation}
\label{cond} 
t-L\leq\vert\mathbf{r}-\mathbf{r}_o\vert/c\leq t,
\end{equation}
where $\vert\mathbf{r}-\mathbf{r}_o\vert$ is the distance of the emitter from the Earth, located at the vector position $\mathbf{r}_o$
(with $\vert\mathbf{r}_o\vert\simeq 27$ kly),
and $c$ is the speed of light. We note that in the case of isotropic technoemissions, Eq.~\eqref{cond} is equivalent to requiring that at present time
the Earth is within a spherical shell region of outer radius $ct$ and thickness $cL$ filled by the emitted electromagnetic 
radiation \citep{Grimaldi2017,Grimaldi2018}.

The causal constraint of Eq.~\eqref{cond} introduces a selection effect for the observable technoemissions, which can profoundly
alter the expected longevity of a detected signals relative to the expected $L$ coming from the underlying distribution $\rho_L(L)$.
Indeed, as pointed out in \citet{Balbi2021}, from Eq.~\eqref{cond} it follows that only technosignals of longevity greater 
than $t-\vert\mathbf{r}_o-\mathbf{r}\vert/c$ are 
detectable in principle, which leads to two important consequences. First, it introduces a correlation between the otherwise 
statistically independent variables $L$ and $t$ and, second, it filters out those signals 
that are too short-lived to cross our planet, depending on the emitter distance and the time of emission. The very event of detection,
therefore, promotes long-lived techoemissions over short-lasting ones, as quantitatively confirmed by Monte Carlo simulations in \citet{Balbi2021}
and by a Bayesian analysis in \citet{Kipping2020}, thereby implying that the first contact will likely be with an older technology than our own. 

As remarked in the Introduction, however, there are arguments suggesting that energy and cost constraints and expected lifetimes of technologies might actually make long-lived technoemissions \textit{a priori} less likely than short-lived ones, leading one to wonder what would then be the longevity of a detected signal and the age of its emitter. To address this issue, we assume that a search for technosignatures has been successful in detecting a signal coming from an emitter within a distance $R_o$ from Earth. Although $R_o$ depends on both the technoemission characteristics and detector 
specifications \citep{Grimaldi2018}, in the following we treat it as a free parameter representing the maximum distance the hypothetical 
search is able to sample. 

We follow \citet{Balbi2021} and equate the probability of 
detecting an emission of longevity $L$ generated at time $t$ with the joint probability that the technoemission fullfils the causal 
condition \eqref{cond} and that $\vert\mathbf{r}-\mathbf{r}_o\vert\leq R_o$:  
\begin{equation}
\label{Prob1}
P(t,L;\mathcal{D})=\int\!d\mathbf{r}\rho_E(\mathbf{r})\theta(R_o-\vert\mathbf{r}-\mathbf{r}_o\vert)f_{t,L}(\mathbf{r}-\mathbf{r}_o),
\end{equation}
where $f_{t,L}(\mathbf{r}-\mathbf{r}_o)=1$ if Eq.~\eqref{cond} is satisfied and $f_{t,L}(\mathbf{r}-\mathbf{r}_o)=0$ otherwise, 
$\theta(x)$ is the Heaviside step function, and $\mathcal{D}$ indicates the event of detection, whose probability of occurrence 
is obtained by marginalizing Eq.~\eqref{Prob1}
over the underlying distributions $\rho_L(L)$ and $\rho_t(t)=\Gamma(t)/\bar{N}$:
 \begin{equation}
 \label{Prob2}
P(\mathcal{D})=\int\!dt\rho_t(t)\int\!dL\rho_L(L)P(t,L;\mathcal{D}).
\end{equation}
Finally, partial marginalization of Eq.~\eqref{Prob1} with respect to $t$ and $L$ and the use of Bayes' theorem allow us to write the
PDFs of, respectively, $L$ and $t$, conditional on the detection:
\begin{equation}
\label{PDFL}
\rho_L(L\vert \mathcal{D})=\frac{\rho_L(L)\int\!dt\rho_t(t) P(t,L;\mathcal{D})}{P(\mathcal{D})},
\end{equation}
\begin{equation}
\label{PDFt}
\rho_t(t\vert \mathcal{D})=\frac{\rho_t(t)\int\!dL\rho_L(L)  P(t,L;\mathcal{D})}{P(\mathcal{D})}.
\end{equation}
Note that the above equations can be viewed as a Bayesian inference of $L$ and $t$, where $\rho_L(L)$ and $\rho_t(t)$
are identifiable as the prior PDFs from which the posteriors $\rho_L(L\vert \mathcal{D})$ and $\rho_t(t\vert \mathcal{D})$ are
inferred, given the event of detection.

\begin{figure*}[t]
\begin{center}
\includegraphics[width=18 cm,clip=true]{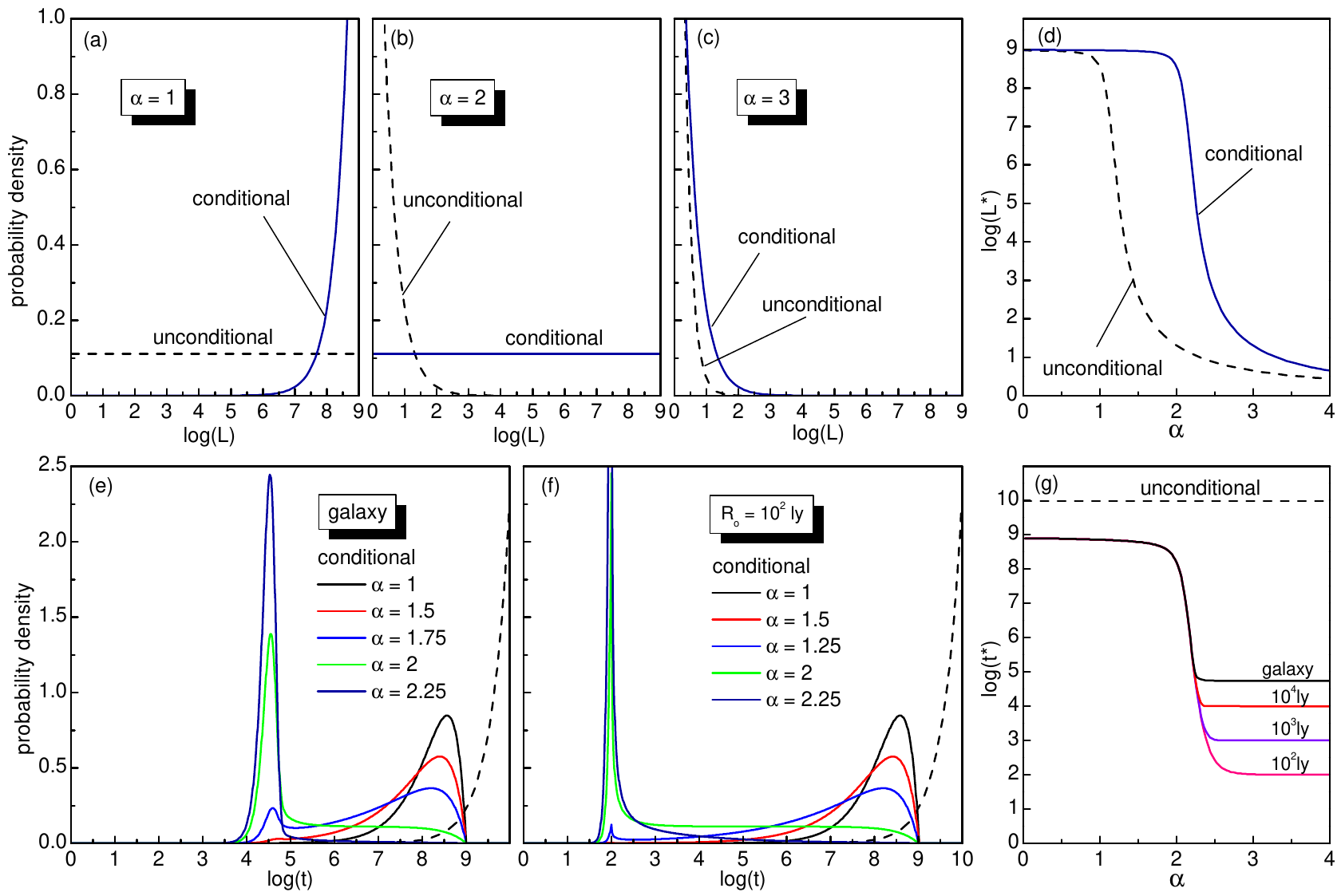}
\caption{PDF of technoemission longevity $L$ (top row) and starting time $t$ of 
emissions (bottom row) calculated under the assumption that the unconditional PDF of $t$ is uniform over the interval $[1,10^{10}]$ yr. 
Panels (a)-(c) show the conditional, Eq.~\eqref{PDFLuni} and unconditional, Eq.~\eqref{power}, PDFs of $L$ for three values of the 
exponent $\alpha$, whereas panel (d) shows the corresponding values of $L^*$ such that $L\leq L^*$ with 95 \% probability.
Panels (e) and (f) show the PDF of $t$ calculated numerically for different values of $\alpha$ and (e) for $R_o$ large enough 
to encompass the entire galaxy and (f) for $R_o=100$ ly. Panel (g) shows the values of $t^*$ such that $t\leq t^*$ with
95 \% probability calculated for different values of $R_o$.}
\label{fig1}
\end{center}
\end{figure*}

As mentioned in the Introduction, we take the underlying PDF of the longevity to be shaped by the Lindy effect 
and adopt for $\rho_L(L)$ the following power-law distribution:
\begin{equation}
\label{power}
\rho_L(L)=\left\{
\begin{array}{ll}
\lambda L^{-\alpha} & L_\textrm{min}\leq L\leq L_\textrm{max}\\
0  & \textrm{otherwise}
\end{array}\right.
\end{equation}
where  $\lambda=(1-\alpha)/(L_\textrm{max}^{1-\alpha}-L_\textrm{min}^{1-\alpha})$ is a normalization constant,
and $L_\textrm{min}=1$ yr and $L_\textrm{max}=10^9$ yr are lower and upper limits on $L$ which we introduce to make $\rho_L(L)$ 
normalizable to unity for any choice of $\alpha$. 
Recall that $\alpha>1$ implies a scenario in which small values of $L$ are favored over large ones, as should be expected if the
longevity of technoemissions is constrained by energy demands.  In the following, however, we will treat the exponent $\alpha$ as a 
free, positive parameter to illustrate its effect on the longevity distribution inferred from a detection event.

\begin{figure*}[t]
\begin{center}
\includegraphics[width=18 cm,clip=true]{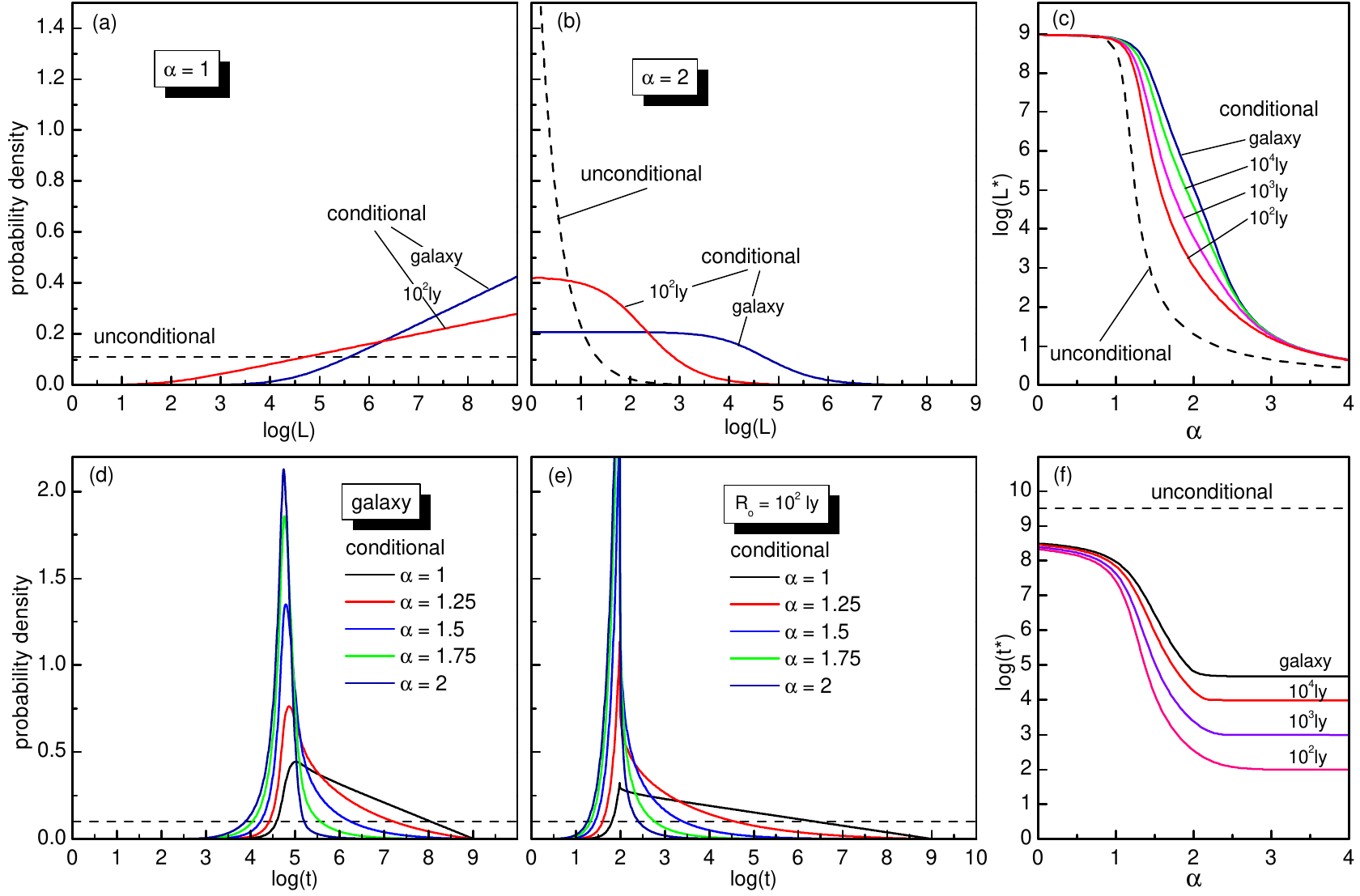}
\caption{PDFs of $L$ (top row) and $t$ of (bottom row) calculated numerically using an unconditional distribution of emission 
starting times that is log-uniform over the interval $[1,10^{10}]$ yr. 
In panels (a) and (b) the conditional PDF of $L$ (solid lines) is plotted different $\alpha$ and $R_o$ values. Panel (c)
shows the resulting $L^*$ for several observation radii $R_o$.
Panels (d) and (e) show the PDF of $t$ calculated numerically for different values of $\alpha$ and (d) for $R_o$ large and (e) 
for $R_o=100$ ly. Panel (f) shows the values of $t^*$ calculated for different values of $R_o$.}
\label{fig2}
\end{center}
\end{figure*}

\section{Results}
\label{results}

We start our analysis by first exploring the scenario in which $\rho_t(t)$ is uniform in $t$ ($\rho_t(t)=\rho_t$), since in this case 
$\rho_L(L\vert R_0)$ can be calculated analytically.  The integration over $t$ appearing in Eqs.~\eqref{Prob2}
and \eqref{PDFL} can indeed be performed exactly, yielding:
\begin{equation}
\label{PDFLuni}
\rho_L(L\vert R_o)=\frac{\rho_L(L)L}{\langle L\rangle}=\frac{(2-\alpha)L^{1-\alpha}}{L_\textrm{max}^{2-\alpha}-L_\textrm{min}^{2-\alpha}},
\end{equation}
where $\langle L\rangle=\int\!dL L\rho_L(L)$ is the unconditional average longevity and
the second equality stems from using for $\rho_L(L)$ the power-law form of Eq.~\eqref{power}. It is apparent from Eq.~\eqref{PDFLuni} that
the probability of detecting a technoemission is proportional to its longevity,
regardless of the observational radius $R_o$ and the spatial distribution of the 
emitters. This is a consequence of using a stationary emission rate, whereas, as shown in the following,
a log-uniform $\rho_t(t)$ does not yield such simple scaling.

As a consequence of  $\rho_L(L\vert R_o)\propto L\rho_L(L)$, the event of detection can entail a significant weight 
shift of $\rho_L(L\vert R_o)$ from small to  large $L$ compared to the unconditional PDF, as shown 
in Figs.~\eqref{fig1}(a)-(b) for $\alpha=1$ and $\alpha=2$.
However, for $\alpha = 3$ (Fig.~\ref{fig1}(c)) the unconditional PDF becomes so skewed toward
small values of $L$ that the causal constraint has little effect on $\rho_L(L\vert R_o)$. Borrowing a Bayesian terminology, we can say that 
for sufficiently large $\alpha$ the prior $\rho_L(L)$ is so informative that completely determines the posterior $\rho_L(L\vert R_o)$.

To see more quantitatively how $\alpha$ affects the longevity of a detected signal, we introduce $L^*$ 
defined as the longevity such that $L\leq L^*$ with $95$ \% probability. 
For a detected signal, this is obtained by requiring that the integral of Eq.~\eqref{PDFLuni} 
in the interval $[L_\textrm{min},L^*]$ be equal to $0.95$, which by isolating $L^*$ yields:
\begin{equation}
\label{Lstar1}
L^*=\left[L_\textrm{min}^{2-\alpha}+0.95(L_\textrm{max}^{2-\alpha}-L_\textrm{min}^{2-\alpha})\right]^{\frac{1}{2-\alpha}}.
\end{equation} 
The formula for the unconditional $L^*$ is obtained by replacing $2-\alpha$ with $1-\alpha$ in Eq.~\eqref{Lstar1}.
Figure \ref{fig1}(d) shows that for $\alpha\lesssim 1$ both the unconditional and conditional $L^*$ are equally large ($\sim 10^9$ yr),
whereas in the interval $1\lesssim\alpha\lesssim 2$ the conditional $L^*$ exceeds by several orders of magnitude the rapidly decreasing 
unconditional $L^*$. It is in this range of $\alpha$ that the event of detection has its strongest selection effect. Finally, upon an 
increase of $\alpha$ beyond $\alpha=2$, the conditional $L^*$ rapidly drops toward $L_\textrm{min}=1$ yr, following asymptotically 
$L^*=20^{1/\alpha} L_\textrm{min}$.

At this point, it is worth noting that Lindy's law with $\alpha>2$ arises from assuming that the future lifetime expectancy
of technoemissions is proportional to their current age (see the Appendix). Should the duration of technoemissions meet this requirement,
 therefore, we should not expect first contact to occur with very long-lived signals.
For example, already for $\alpha=3$, we find from Eq.~\eqref{Lstar1} that the longevity of a detected signal is smaller than $20$ yr, with
$95$ \% probability (to be compared to a unconditional longevity smaller than about $4.5$ yr). However, as shown by \citet{Eliazar2017},
an alternative mathematical definition of the Lindy effect can be formulated through the median (instead of the average) future lifespan 
being proportional to the current age.
In this case, the power-law distribution of $L$ would have an exponent $\alpha>1$. 

We turn now our attention to the conditional PDF of $t$ and its evolution with $\alpha$, shown in Fig.~\ref{fig1}(e) for $R_o$ large
enough to encompass the entire galaxy and in Fig.~\ref{fig1}(f) for $R_o$ of only $100$ ly. Two features are apparent. 
First, for $\alpha=1$ and regardless of the observation distance $R_o$, the starting time of a detected signal is broadly distributed 
around $3\times 10^8$ yr, as found by the Monte Carlo simulations of \citet{Balbi2021}. This means that among the technoemissions 
that do not have preferred scale on $L$, those that can be detected likely have very large longevities and were generated 
several millions of years ago.
The second result is that by increasing $\alpha$, the age $t$ of a detectable signal can be considerably reduced, as the weight of 
$\rho_t(t\vert \mathcal{D})$ is gradually transferred to much smaller times, eventually forming a single peak centered around 
$3\times 10^4$ yr for $R_o$ large and $\sim 10^2$ yr for $R_o=100$ ly.
This is understood by recognizing that for $\alpha > 2$ the technoemissions are so short lived (as shown by the dashed line in Fig.~\ref{fig1}(c)) that
the causal constraint of Eq.~\eqref{cond} selects only values of $t$ near $\vert \mathbf{r}-\mathbf{r}_o\vert /c$, whose most probable value for the emitter distribution of Eq.~\eqref{rhoE} is at $\min(3\times 10^4 \textrm{yr}, R_o/c)$.

The starting emission times of a detectable signal are summarized in Fig~\ref{fig1}(g), where we show the computed time $t^*$ such that
$t\leq t^*$ with $95$ \% probability. Two regimes are evident: for $\alpha\lesssim 2$, where the longevity is expected to be very large,
$t^*$ conditioned to the event of detection is about $10^9$ yr, whereas for $\alpha\gtrsim 2$ the conditional $t^*$ rapidly drops by 
several orders of magnitude toward times that depend on the sampled distance $R_o$. From the evolution of $L^*$ and $t^*$ with $\alpha$
shown in Figs.~\ref{fig1}(d) and \ref{fig1}(g), we conclude therefore that the event of detection \textit{per se} does not necessarily imply that
first contact will be with very long lasting and very old technoemissions, and that relatively young and short-lived signals have the highest 
probability of being detected once $\alpha\gtrsim 2$. 

The assumption of a stationary emission rate (that is, a constant $\rho_t(t)$) is questionable in several respects, not least the fact that
the habitability of the galaxy is itself a function of time. Adopting a PDF of $t$ that reproduces the main feature of the temporal dependence
of the galactic habitable zone of \citet{Lineweaver2004}  (i.e., the age of planets suitable to host complex life is normally distributed, with mean $5.5$ Gyr and standard deviation of $2$ Gyr, with a fiducial interval of $4$ Gyr for the appearance
of a technological species; see also \citet{Balbi2021}) essentially replicates 
the results of Fig.~\ref{fig1}, whereas assuming a log-uniform PDF of $t$ leads to some quantitative changes but does not substantially alter
the conclusions reported above. First, in contrast with the case of a constant $\rho_t(t)$, the conditional PDF of the longevity depends on the
sampled distance $R_o$, as summarized in Figs.~\ref{fig2}(a)-(c). In addition, we find that the selection effect of Eq.~\eqref{cond} is less
effective in promoting large conditional longevities, as seen by the $L^*$ versus $\alpha$ plot of Fig.~\ref{fig2}(c). Second, compared
to the case $\rho_t(t)=\textrm{constant}$, an increase of $\alpha$ entails a more rapid weight shift of the conditional PDF of $t$ toward 
$\min(3\times 10^4 \textrm{yr}, R_o/c)$, as shown in Fig.~\ref{fig2}(d)-(f).

\begin{figure*}[t]
\begin{center}
\includegraphics[width=18 cm,clip=true]{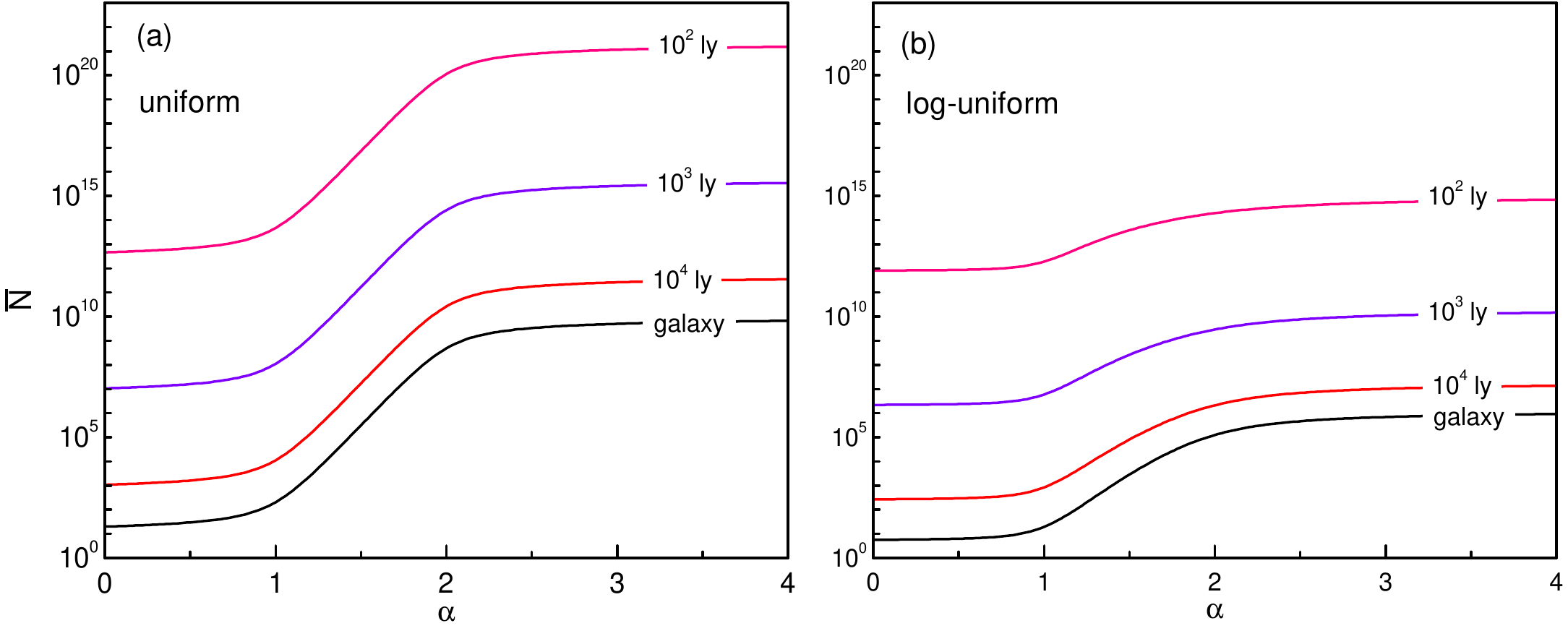}
\caption{Number of technoemissions ever generated over the history of the Galaxy as a function of the longevity exponent $\alpha$. 
$\bar{N}$ is calculated from Eq.~\eqref{Nbar} using a uniform in $t$ (a) and a log-uniform in $t$ (b) probability disribution
of the time of occurrence of technoemissions. Results are shown for different values of the sampled radius $R_o$.} 
\label{fig3}
\end{center}
\end{figure*}

\section{Discussion and conclusions}
\label{concl}

The main conclusion of our analysis is that assuming different power-law shapes for the probability distribution of technosignature longevities has a sizable effect on the most likely duration of detected technosignatures. We confirmed previous findings that showed that the expected longevity of detected technosignatures will generally be very large \citep{Kipping2020, Balbi2021}, but we showed that this is strictly true only for rather uniformative priors, with $\alpha$ not too different from 1. In particular, we found that in the regime $1\lesssim\alpha \lesssim 2$ the most likely technoemissions to be detected have  $L \sim 10^9$ yr (with $95 \%$ confidence). 

However, when we make the reasonable assumption that short-lived technoemissions vastly outnumber the long-lived ones (as it is the case if their operation has an energy or maintenance cost that increases with time), then the first to be detected will likely have a relatively short $L$. This starts to become apparent for power-law distributions with $\alpha\gtrsim 2$ (characteristic of a Lindy effect behavior); for $\alpha > 3$ the preference for large $L$ essentially disappears, with the most likely to be detected having $L$ much smaller than $\sim 10^2$ yr (at $95 \%$ confidence; Fig. \ref{fig1}(d)). This is a significant departure from previous studies. 

The selection effect is further reduced if we assume that the epoch of appearance of technosignatures over the history of the Galaxy has a log-uniform (rather than uniform) distribution in $t$. In this case, already in the regime $\alpha \gtrsim 1$ the most likely detected longevity is well below $\sim 10^9$ yr, regardless of the radius of the surveyed volume (Fig. \ref{fig2}c). 

\begin{figure}[t]
\begin{center}
\includegraphics[width=8.5 cm,clip=true]{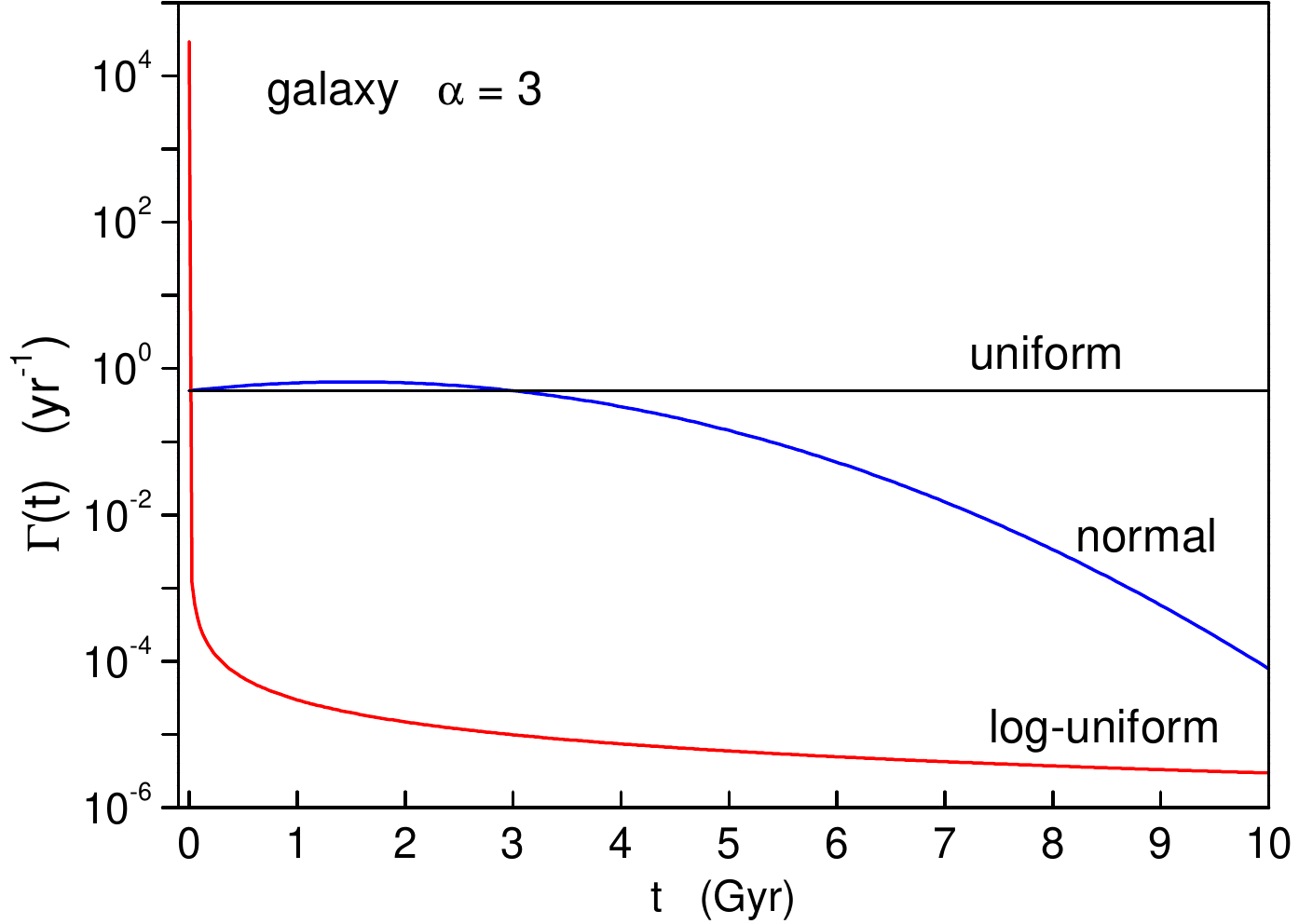}
\caption{Technoemission rates calculated from $\Gamma(t)=\bar{N}\rho_t(t)$ for uniform, log-uniform, and normally distributed PDFs
of $t$. For all cases, $\bar{N}$ is calculated for the entire galaxy and for $\alpha=3$.}
\label{fig4}
\end{center}
\end{figure}

A final point is worth discussing; namely, what is the number of emitters $\bar N$ that had to appear, over the history of the Galaxy, in order that, typically, one of them satisfies the condition of Eq. (\ref{cond}), and is therefore detectable today from Earth? 
The answer, again, depends on the assumed distribution of $L$, and is obtained by requiring that $\bar{N}P(\mathcal{D})=1$, where
$P(\mathcal{D})$ is the detection probability given in Eq.~\eqref{Prob2}. Figure \ref{fig3} shows
\begin{equation}
\label{Nbar}
\bar{N}=\frac{1}{P(\mathcal{D})}
\end{equation}
calculated as a function of the exponent $\alpha$ for uniform in $t$ (Fig. \ref{fig3}(a)) and log-uniform in $t$ (Fig. \ref{fig3}(b)) distributions of the unconditional starting times of emission. 
In both cases, $\bar{N}$ increases by several orders of magnitude as $\alpha$ goes from 
$\alpha\lesssim 1$, where it takes only a few long-lived technoemissions to have typically one detection, to $\alpha\gtrsim 2$,
where there must be many short-lived ones for one to be detectable. In particular, if the epoch of appearance is uniformly distributed 
over the history of the Galaxy (Fig. \ref{fig3}(a)), then $\bar N\approx 10^{10}$ for $\alpha\gtrsim 2$ and $R_o$
such that the entire Milky Way is sampled. This would imply a constant rate of about one technoemission per year generated over the last $\approx 10$ Gyr, as seen in Fig.~\ref{fig4} for $\alpha=3$, where we show that a similar rate would also occur during 
the last $\approx 3$ Gyr if the PDF of $t$ followed a normal distribution. Although high, rates of this magnitude do not exceed those
of some natural astrophysical phenomena, such as star formation and novae in the Milky Way. 
For a log-uniform emission rate, Fig.~\ref{fig3}(b), we find that $\bar{N}$ can be reduced by several orders of magnitude compared to the previous scenarios. However, since in this case $\rho_t(t)\propto 1/t$, the rate would be such that $20$ \% of the total 
number of emissions ever produced, $\bar{N}\approx 5\times 10^5$, must be generated only in the last $100$ yr.

Reducing the sampled volume inevitably increases $\bar{N}$, as seen in Fig.~\ref{fig3} for different values of $R_o$, leading to 
exceedingly high emission rates. This is because $P(\mathcal{D})$ becomes small for reduced $R_o$, thus leading
$\bar{N}$ to increase in order to have one detectable technoemission, especially if $\alpha$ is large. It is worth pointing out, however, that if the emitters were less uniformly distributed than the density profile of Eq.~\eqref{rhoE}, different 
estimates of $\bar{N}$ would then be expected. Assuming for example that our planet belongs to a more or less localized cluster 
of planets harboring technological species, then $\bar{N}$ could be greatly reduced, in analogy to what would happen to 
the number of biospheres in a panspermia scenario \citep{Balbi2020}.


\appendix
\section{Lindy's law}
\label{Lindy}
Lindy's law derives from the empirical observation that the future lifetime of non-perishable things such as technology 
increases with its current age. Here we show that if we assume that the future life expectancy of technoemissions is 
proportional to their current age, then the PDF of their longevities follows the power-law of Eq.\eqref{PDFL} with $\alpha>2$. 
In the following derivation, we adapt the formulation of \citet{Eliazar2017} to our notation. 

We denote $\rho_L(L)$ the unconditional PDF of the longevity, which we treat as a random variable $X_L$ defined 
in the interval $[L_\textrm{min},L_\textrm{max}]$.  The corresponding complementary cumulative distribution function (also known as 
survival function) is $P_L(L)=\int_L^\infty\!ds \rho_L(s)$, from which we have also that $\rho_L(L)=-P_L'(L)$. In terms of the random
variable $X_L$, Lindy's law can be formulated as:
\begin{equation}
\label{Lindy1}
\mathbf{E}[X_L-L\vert X_L\geq L]=aL
\end{equation}
where $a$ is a positive constant of proportionality. The left-hand side of \eqref{Lindy1} can be written in terms of $\rho_L$ and $P_L$ as follows:
\begin{eqnarray}
\label{Lindy2}
\mathbf{E}[X_L-L\vert X_L\geq L]&=&\dfrac{\int_L^\infty\!ds (s-L)\rho_L(s)}{\int_L^\infty\!ds \rho_L(s)} \nonumber \\
&=&-\dfrac{\int_L^\infty\!\!ds\,s P_L'(s)}{P_L(L)}-L.
\end{eqnarray}
Equating the right-hand sides of Eqs. \eqref{Lindy1} and \eqref{Lindy2}, and derivating the resulting equation with respect to $L$ we obtain: 
\begin{equation}
LP_L'(L)=(1+a)P_L(L)+(1+a)LP_L^(L),
\end{equation}
which can be rearranged to obtain the following differential equation for $P_L$:
\begin{equation}
\frac{P_L'(L)}{P_L(L)}=-\frac{1+a}{a}\frac{1}{L},
\end{equation}
whose solution is $P_L(L)\propto L^{-(1+1/a)}$, and hence using $\rho_L(L)=-P_L'(L)$:
\begin{equation}
\rho_L(L)\propto L^{-\alpha},
\end{equation}
where $\alpha=2+1/a$. Since $a>0$, we obtain therefore $\alpha>2$.

\bibliographystyle{aasjournal}
\bibliography{Longevity2}

\begin{thebibliography}{}
\expandafter\ifx\csname natexlab\endcsname\relax\def\natexlab#1{#1}\fi
\providecommand{\url}[1]{\href{#1}{#1}}
\providecommand{\dodoi}[1]{doi:~\href{http://doi.org/#1}{\nolinkurl{#1}}}
\providecommand{\doeprint}[1]{\href{http://ascl.net/#1}{\nolinkurl{http://ascl.net/#1}}}
\providecommand{\doarXiv}[1]{\href{https://arxiv.org/abs/#1}{\nolinkurl{https://arxiv.org/abs/#1}}}

\bibitem[{Balbi(2018)}]{Balbi2018}
Balbi, A. 2018, Astrobiology, 18, 54, \dodoi{10.1089/ast.2017.1652}

\bibitem[{Balbi \& Grimaldi(2020)}]{Balbi2020}
Balbi, A., \& Grimaldi, C. 2020, Proceeding of the National Academy of
  Sciences, 117, 21031, \dodoi{10.1073/pnas.2007560117}

\bibitem[{Balbi \& Ćirković(2021)}]{Balbi2021}
Balbi, A., \& Ćirković, M.~M. 2021, The Astronomical Journal, 161, 222,
  \dodoi{10.3847/1538-3881/abec48}

\bibitem[{Benford {et~al.}(2010)Benford, Benford, \& Benford}]{Benford2010a}
Benford, J., Benford, G., \& Benford, D. 2010, Astrobiology, 10, 475,
  \dodoi{10.1089/ast.2009.0393}

\bibitem[{\'Cirkovi\'c {et~al.}(2019)\'Cirkovi\'c, Vukoti\'c, \&
  Stojanovi\'c}]{Cirkovic2019}
\'Cirkovi\'c, M.~M., Vukoti\'c, B., \& Stojanovi\'c, M. 2019, Astrobiology, 19,
  1300, \dodoi{10.1089/ast.2019.2052}

\bibitem[{Davies \& Wagner(2013)}]{Davies2013}
Davies, P. C.~V., \& Wagner, R.~V. 2013, Acta Astronautica, 89, 261,
  \dodoi{10.1016/j.actaastro.2011.10.022}

\bibitem[{Dyson(1960)}]{Dyson1960}
Dyson, F.~J. 1960, Science (New York, N.Y.), 131, 1667,
  \dodoi{10.1126/science.131.3414.1667}

\bibitem[{Eliazar(2017)}]{Eliazar2017}
Eliazar, I. 2017, Physica A: Statistical Mechanics and its Applications, 486,
  797, \dodoi{https://doi.org/10.1016/j.physa.2017.05.077}

\bibitem[{Forgan \& Elvis(2011)}]{Forgan2011}
Forgan, D.~H., \& Elvis, M. 2011, International Journal of Astrobiology, 10,
  307–313, \dodoi{10.1017/S1473550411000127}

\bibitem[{Gray(2020)}]{Gray2020}
Gray, R.~H. 2020, International Journal of Astrobiology, 19, 299,
  \dodoi{10.1017/S1473550420000038}

\bibitem[{Grimaldi(2017)}]{Grimaldi2017}
Grimaldi, C. 2017, Scientific Reports, 7, 46273, \dodoi{10.1038/srep46273}

\bibitem[{Grimaldi \& Marcy(2018)}]{Grimaldi2018}
Grimaldi, C., \& Marcy, G.~W. 2018, Proceedings of the National Academy of
  Sciences, 115, E9755, \dodoi{10.1073/pnas.1808578115}

\bibitem[{Haqq-Misra {et~al.}(2022{\natexlab{a}})Haqq-Misra, Fauchez,
  Schwieterman, \& Kopparapu}]{Haqq-Misra2022a}
Haqq-Misra, J., Fauchez, T.~J., Schwieterman, E.~W., \& Kopparapu, R.
  2022{\natexlab{a}}, The Astrophysical Journal Letters, 929, L28,
  \dodoi{10.3847/2041-8213/ac65ff}

\bibitem[{Haqq-Misra \& Kopparapu(2012)}]{Haqq-Misra2012}
Haqq-Misra, J., \& Kopparapu, R. 2012, Acta Astronautica, 72, 15,
  \dodoi{10.1016/j.actaastro.2011.10.010}

\bibitem[{Haqq-Misra {et~al.}(2022{\natexlab{b}})Haqq-Misra, Kopparapu,
  Fauchez, Frank, Wright, \& Lingam}]{Haqq-Misra2022b}
Haqq-Misra, J., Kopparapu, R., Fauchez, T.~J., {et~al.} 2022{\natexlab{b}}, The
  Planetary Science Journal, 3, 60, \dodoi{10.3847/PSJ/ac5404}

\bibitem[{Kipping {et~al.}(2020)Kipping, Frank, \& Scharf}]{Kipping2020}
Kipping, D., Frank, A., \& Scharf, C. 2020, International Journal of
  Astrobiology, 1, \dodoi{10.1017/S1473550420000208}

\bibitem[{Kopparapu {et~al.}(2021)Kopparapu, Arney, Haqq-Misra, Lustig-Yaeger,
  \& Villanueva}]{Kopparapu2021}
Kopparapu, R., Arney, G., Haqq-Misra, J., Lustig-Yaeger, J., \& Villanueva, G.
  2021, The Astrophysical Journal, 908, 164, \dodoi{10.3847/1538-4357/abd7f7}

\bibitem[{Lin {et~al.}(2014)Lin, Abad, \& Loeb}]{Lin2014}
Lin, H.~W., Abad, G.~G., \& Loeb, A. 2014, The Astrophysical Journal, 792, L7,
  \dodoi{10.1088/2041-8205/792/1/L7}

\bibitem[{Lineweaver {et~al.}(2004)Lineweaver, Fenner, \&
  Gibson}]{Lineweaver2004}
Lineweaver, C.~H., Fenner, Y., \& Gibson, B.~K. 2004, Science, 303, 59,
  \dodoi{10.1126/science.1092322}

\bibitem[{Lingam \& Loeb(2017{\natexlab{a}})}]{Lingam2017}
Lingam, M., \& Loeb, A. 2017{\natexlab{a}}, Monthly Notices of the Royal
  Astronomical Society: Letters, 470, L82, \dodoi{10.1093/mnrasl/slx084}

\bibitem[{Lingam \& Loeb(2017{\natexlab{b}})}]{Lingam2017a}
---. 2017{\natexlab{b}}, The Astrophysical Journal, 837, L23,
  \dodoi{10.3847/2041-8213/aa633e}

\bibitem[{{Mandelbrot}(1984)}]{Mandelbrot1984}
{Mandelbrot}, B.~B. 1984, {The Fractal Geometry of Nature} (San Francisco, CA:
  Freeman)

\bibitem[{Margot {et~al.}(2019)Margot, Croft, Lazio, Tarter, \&
  Korpela}]{Margot2019}
Margot, J.-L., Croft, S., Lazio, J., Tarter, J., \& Korpela, E. 2019, Bulletin
  of the AAS, 51

\bibitem[{Newman(2005)}]{Newman2005}
Newman, M. 2005, Contemporary Physics, 46, 323,
  \dodoi{10.1080/00107510500052444}

\bibitem[{Ord(2023)}]{ord2023lindy}
Ord, T. 2023, The Lindy Effect.
\newblock \doarXiv{2308.09045}

\bibitem[{Schwartz \& Townes(1961)}]{Schwartz1961}
Schwartz, R.~N., \& Townes, C.~H. 1961, Nature, 190, 205,
  \dodoi{10.1038/190205a0}

\bibitem[{Socas-Navarro(2018)}]{Socas-Navarro2018}
Socas-Navarro, H. 2018, The Astrophysical Journal, 855, 110,
  \dodoi{10.3847/1538-4357/aaae66}

\bibitem[{Socas-Navarro {et~al.}(2021)Socas-Navarro, Haqq-Misra, Wright,
  Kopparapu, Benford, \& Davis}]{Socas-Navarro2021}
Socas-Navarro, H., Haqq-Misra, J., Wright, J.~T., {et~al.} 2021, Acta
  Astronautica, 182, 446, \dodoi{10.1016/j.actaastro.2021.02.029}

\bibitem[{{Taleb}(2012)}]{Taleb2012}
{Taleb}, N.~N. 2012, {Antifragile: Things That Gain from Disorder } (New York:
  Random house)

\bibitem[{Tarter(2006)}]{Tarter2006}
Tarter, J.~C. 2006, Proceedings of the International Astronomical Union, 2, 14,
  \dodoi{10.1017/S1743921307009829}

\bibitem[{Wright(2020)}]{Wright2020}
Wright, J. 2020, Serbian Astronomical Journal, 1, \dodoi{10.2298/SAJ2000001W}

\bibitem[{Wright \& Kipping(2019)}]{Wright2019}
Wright, J., \& Kipping, D. 2019, Bulletin of the AAS, 51

\bibitem[{Wright {et~al.}(2018)Wright, Horowitz, Maire, Werthimer, Antonio,
  Aronson, Chaim-Weismann, Cosens, Drake, Howard, Marcy, Raffanti, Siemion,
  Stone, Treffers, \& Uttamchandani}]{Wright2018}
Wright, S.~A., Horowitz, P., Maire, J., {et~al.} 2018, in Ground-based and
  Airborne Instrumentation for Astronomy VII, ed. C.~J. Evans, L.~Simard, \&
  H.~Takami, Vol. 10702 (SPIE), 107025I, \dodoi{10.1117/12.2314268}

\end{thebibliography}

\end{document}